\documentclass{pasj00}

\begin{document}
\SetRunningHead{Shidatsu et al.}
{Monitoring GX 339$-$4 in the High/soft State with MAXI/GSC}

\title{Long-term Monitoring of the Black Hole Binary GX 339$-$4 in the High/Soft State 
during the 2010 Outburst with MAXI/GSC}

\author{
Megumi \textsc{Shidatsu},\altaffilmark{1}
Yoshihiro \textsc{Ueda},\altaffilmark{1}
Satoshi \textsc{Nakahira},\altaffilmark{2}
Hitoshi \textsc{Negoro},\altaffilmark{3}
Kazutaka \textsc{Yamaoka},\altaffilmark{4}
Mutsumi \textsc{Sugizaki},\altaffilmark{2}
Kazuo \textsc{Hiroi},\altaffilmark{1}
Nobuyuki \textsc{Kawai},\altaffilmark{5}
Tatehiro \textsc{Mihara},\altaffilmark{2}
Masaru \textsc{Matsuoka},\altaffilmark{2,6}
Masashi \textsc{Kimura},\altaffilmark{7}
Masaki \textsc{Ishikawa},\altaffilmark{8}
Naoki \textsc{Isobe},\altaffilmark{9}
Hiroki \textsc{Kitayama},\altaffilmark{7}
Mitsuhiro \textsc{Kohama},\altaffilmark{6}
Takanori \textsc{Matsumura},\altaffilmark{10}
Mikio \textsc{Morii},\altaffilmark{5}
Yujin E. \textsc{Nakagawa},\altaffilmark{11}
Motoki \textsc{Nakajima},\altaffilmark{12}
Motoko \textsc{Serino},\altaffilmark{2}
Tetsuya \textsc{Sootome},\altaffilmark{2,13}
Kousuke \textsc{Sugimori},\altaffilmark{5}
Fumitoshi \textsc{Suwa},\altaffilmark{3}
Takahiro \textsc{Toizumi},\altaffilmark{5}
Hiroshi \textsc{Tomida},\altaffilmark{6}
Yohko \textsc{Tsuboi},\altaffilmark{10}
Hiroshi \textsc{Tsunemi},\altaffilmark{7}
Shiro \textsc{Ueno},\altaffilmark{6}
Ryuichi \textsc{Usui},\altaffilmark{5}
Takayuki \textsc{Yamamoto},\altaffilmark{2,3}
Kyohei \textsc{Yamazaki},\altaffilmark{10}
and Atsumasa \textsc{Yoshida},\altaffilmark{4}
}
\altaffiltext{1}{Department of Astronomy, Kyoto University, Oiwake-cho, Sakyo-ku, Kyoto 606-8502}
\email{shidatsu@kusastro.kyoto-u.ac.jp}
\altaffiltext{2}{MAXI team, Institute of Physical and Chemical Research (RIKEN), 2-1 Hirosawa, Wako, Saitama 351-0198}
\altaffiltext{3}{Department of Physics, Nihon University, 1-8-14 Kanda-Surugadai, Chiyoda-ku, Tokyo 101-8308}
\altaffiltext{4}{Department of Physics and Mathematics, Aoyama Gakuin University,\\ 
5-10-1 Fuchinobe, Chuo-ku, Sagamihara, Kanagawa 252-5258}
\altaffiltext{5}{Department of Physics, Tokyo Institute of Technology, 2-12-1 Ookayama, Meguro-ku, Tokyo 152-8551}
\altaffiltext{6}{ISS Science Project Office, Institute of Space and Astronautical Science (ISAS), 
Japan Aerospace Exploration Agency (JAXA), 2-1-1 Sengen, Tsukuba, Ibaraki 305-8505}
\altaffiltext{7}{Department of Earth and Space Science, Osaka University, 1-1 Machikaneyama, Toyonaka, Osaka 560-0043}
\altaffiltext{8}{School of Physical Science, Space and Astronautical Science, The graduate University 
for Advanced Studies (Sokendai), Yoshinodai 3-1-1, Chuo-ku, Sagamihara, Kanagawa 252-5210}
\altaffiltext{9}{Institute of Space and Astronautical Science (ISAS), Japan Aerospace Exploration Agency (JAXA),
3-1-1 Yoshino-dai, Chuo-ku, Sagamihara, Kanagawa 252-5210}
\altaffiltext{10}{Department of Physics, Chuo University, 1-13-27 Kasuga, Bunkyo-ku, Tokyo 112-8551}
\altaffiltext{11}{Research Institute for Science and Engineering, Waseda University, 17 Kikui-cho, Shinjuku-ku, Tokyo 162-0044}
\altaffiltext{12}{School of Dentistry at Matsudo, Nihon University, 2-870-1 Sakaecho-nishi, Matsudo, Chiba 101-8308}
\altaffiltext{13}{Department of Electronic Information Systems, Shibaura Institute of Technology,\\
307 Fukasaku, Minuma, Saitama, Saitama 337-8570} 


\KeyWords{accretion, accretion disks --- black hole physics --- 
stars: individual(GX 339--4) --- X-rays: binaries} 

\maketitle

\begin{abstract}
We present the results of monitoring the Galactic black hole candidate 
GX 339$-$4 with the Monitor of All-sky X-ray Image (MAXI) / Gas Slit
Camera (GSC) in the high/soft state during the outburst in 2010. All
the spectra throughout the 8-month period are well reproduced with a
model consisting of multi-color disk (MCD) emission and its
Comptonization component, whose fraction is $\leq 25$\% in the total
flux. In spite of the flux variability over a factor of 3, the
innermost disk radius is constant at $R_{\rm in} = 61 \pm 2$ km for
the inclination angle of $i=46^\circ$ and the distance of $d=8$
kpc. 
This $R_{\rm in}$ value is consistent with those of the past
measurements with Tenma in the high/soft state. Assuming that the
disk extends to the innermost stable circular orbit of a non-spinning
black hole, we estimate the black hole mass to be $M=6.8 \pm0.2
M_{\odot}$ for $i=46^\circ$ and $d=8$ kpc, which is
consistent with
that estimated from the Suzaku observation of the previous low/hard
state. Further combined with the mass function, 
we obtain the mass constraint of $4.3 M_\odot <M<
13.3 M_\odot$ for the allowed range of $d=6-15$ kpc and $i<60^\circ$.
We also discuss the spin parameter of the black hole in GX 339--4 by
applying relativistic accretion disk models to the Swift/XRT data.

\end{abstract}

\section{Introduction}

Black hole binaries (BHBs) exhibit various types of X-ray spectra
according to their ``state'', which is mainly determined by the mass
accretion rate (see e.g., \cite{don07} for a recent
review). Typically, at low luminosities, they stay in the ``low/hard
state'', in which the spectrum is well reproduced by a power law with
a photon index of $\Gamma \simeq 1.7$ and an exponential cutoff at
$\sim$100--200 keV. In this state, the accretion disk is likely
truncated (e.g., \cite{shi11}) and its inner part is supposed to be
surrounded by hot corona responsible for Comptonization, although the
interpretation is still in debate (e.g., \cite{mil06}). When a
transient BHB undergoes an outburst, it often makes transition from
the low/hard state to the ``high/soft state'', through the
``intermediate state'', after the rapid increase of the luminosity.

The X-ray spectra of BHBs in the high/soft state are well described by
a Multi-Color Disk (MCD) model \citep{mit84} accompanied with a small
power law with a photon index of $\Gamma \sim$ 2--2.5
(\cite{don07}; \cite{mcc06}). A well accepted picture of this state
is that the optically thick, geometrically thin disk (so-called the
standard disk; \cite{sha73}) is extended down to the innermost stable
circular orbit (ISCO). This is strongly supported by the fact that the
innermost disk radius estimated from the temperature and luminosity of
the MCD model is found to be remarkably constant over a wide
luminosity range for a BHB in the high/soft state (e.g.,
\cite{ebi94}).  Since the ISCO depends only on the black hole mass and
spin, the MCD parameters can give a constraint on the black hole mass
for an assumed spin if the distance and inclination are obtained
(e.g., \cite{dot97} for Cyg X-1), independently from the mass function
derived from the orbital motion.

X-ray monitoring observations of BHBs during their outbursts give
important information on the evolution of the accretion disk as a
function of mass accretion rate. In 2009 September, the Monitor of
All-sky X-ray Image (MAXI; \cite{mat09}) on the International Space
Station (ISS) started its operation. One goal of the MAXI mission is
to detect new transients including BHBs (e.g., \cite{nak10}) and
monitor them as well as known X-ray transients with unprecedented
sensitivities. In fact, MAXI has detected several outbursts of new
BHBs (e.g., MAXI~J1659--152, \cite{neg10a}; MAXI~J1543--564,
\cite{neg10b}) for about 1.5 years. In addition, MAXI gives us a
unique opportunity provides data to study the X-ray spectra of BHBs
continuously over a period from the onset of an outburst to its fading
phase.

GX 339$-$4 is a well known Galactic BHB discovered in early 1970s,
although its binary-system parameters have not been firmly determined
because of the faint magnitude of the companion star. 
\citet{hyn03} estimate the mass function of GX 339--4 to be
$(5.8\pm0.5) M_\odot$ ($2.0 M_\odot$ at a 95\% confidence lower limit)
from high-resolution optical spectroscopy.  \citet{shi11} have
summarized all available constraints on the distance $d$ and
inclination $i$ from various observations. 
The absence of eclipse places a limit of inclination as $i \lesssim
60^\circ$ \citep{cow02}. A tighter constraint, $i = 46^\circ \pm
8^\circ$, is derived by \citet{shi11} from the Suzaku spectra in 2009
March in the low/hard state. The distance is constrained by
\citet{hyn04} to be $6 < d < 15$ kpc from the structure of the Na D
line, and \citet{zdz04} suggest $d=8\pm1$ kpc from the apparent radius
of the secondary low-mass star.

In 2010 April, GX 339--4 underwent an outburst, and a hard-to-soft
transition was observed \citep{mot10}. This source was almost
continuously observed with MAXI during the outburst. In this paper, we
report the first results of the spectral analysis of the MAXI Gas Slit
Camera (GSC; \cite{mih11}) during the high/soft state after the
transition, with a primary purpose to constrain the black hole mass
from the X-ray data. 
We compare the results with those from three
Swift/XRT observations performed simultaneously. 
We also attempt to estimate the spin parameter of the black hole in GX
339--4 by applying relativistic disk models to the Swift/XRT data. 
In Section 2, we describe the observations and data reduction of the
MAXI/GSC and Swift/XRT data. These results are given in Section 3. We
summarize and discuss the results in Section 4. The spectral fit is
performed on XSPEC version 12.6.0q.  The quoted errors refer to 90\%
confidence ranges for a single parameter.

\section{Observation and Data Reduction}
\subsection{Monitoring with MAXI/GSC}

In this paper, we only use the GSC data, which covers the energy band 
of 2--30 keV, considering its wider field-of-view
(FoV) and much larger collecting area than the Solid-state Slit Camera
on MAXI (\cite{tom11}, \cite{tsu10}). The GSC is composed of 12
position-sensitive proportional counters with 6 carbon anodes in each,
although four counters has not been operated since 2010 March due to 
hardware trouble by discharge events.

\begin{figure}
  \begin{center}
    \FigureFile(80mm,80mm){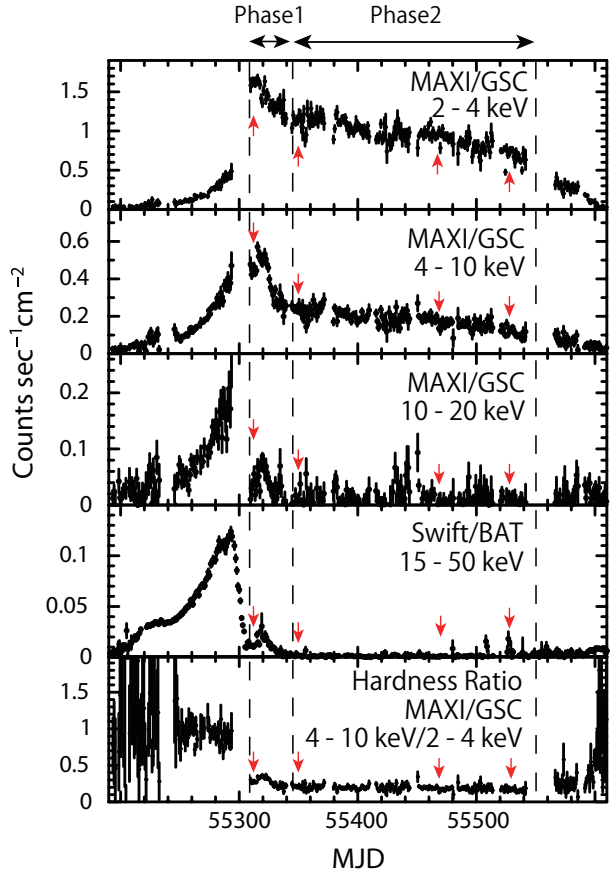}
  \end{center}
  \caption{From top to bottom, light curves of GX 339--4 in the 2--4
    keV, 4--10 keV, and 10--20 keV bands obtained with MAXI/GSC, that
    in the 15--50 keV band obtained with Swift/BAT, and hardness
    ratio between the MAXI/GSC 4--10 keV and 2--4 keV bands. The arrows
    indicate the periods for the sample spectra shown in Figure~\ref{maxi_fit}.}
\label{maxi_publc}
\end{figure}

GX 339$-$4 entered into a new outburst phase in 2010 January
\citep{yam10}, and returned to a quiescent state in 2011 March
\citep{rus11}. This outburst was almost continuously monitored by
MAXI/GSC. Figure~\ref{maxi_publc} shows the GSC light curves of GX
339$-$4 in the energy ranges of 2--4 keV (soft band), 4--10 keV
(medium band), and 10--20 keV (hard band)\footnote{GSC light curves of
about 250 other X-ray sources are available at
http://maxi.riken.jp}. The Swift/BAT light curve in the 15--50
keV band and hardness ratio between the medium and soft bands
of the GSC are also plotted. There are data gap in the MAXI/GSC 
light curves on MJD 55294--55308 because the source was located close to
the direction of the rotation axis of the ISS and was out of the GSC
field-of-views. The total X-ray luminosity most likely reached its peak
during the period. We find that GX 339--4 stayed in the low/hard state 
until MJD 55294, showing a power-law-like spectrum with a photon
index of $\approx$1.7. In this paper, we concentrate on the data taken
during the period between MJD 55310 and 55550, which roughly correspond to 
the flux peak in the total band and the end of the outburst,
respectively. As shown in Figure~\ref{maxi_publc}, the source flux 
in the hard band during MJD 55345--55550 was as small as that before 
the outburst, indicating that it had been stayed in the high/soft state 
during the period.

\subsection{Data Reduction}

\begin{figure}
  \begin{center}
    \FigureFile(80mm,80mm){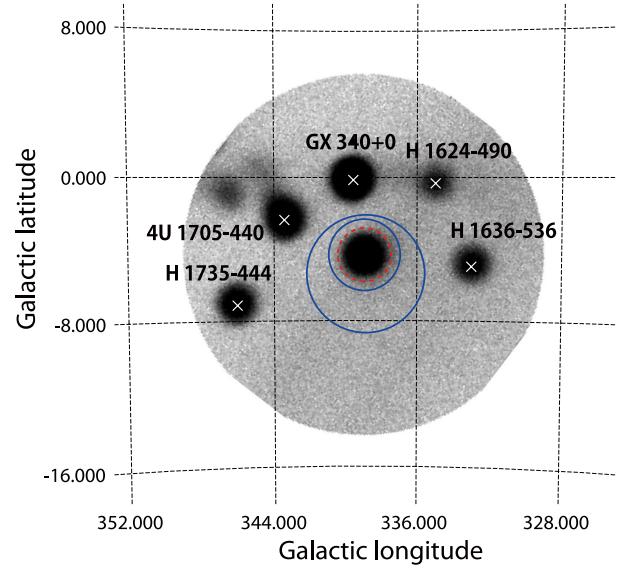}
  \end{center}
  \caption{The MAXI/GSC image integrated from MJD 55350 to MJD 55500.
The event extraction region of the source is plotted with the red
dashed circle, while that of the background corresponds to the area 
between the two blue circles (solid line).}
\label{maxi_img}
\end{figure}

We extract GSC data from the event files 
processed with the MAXI standard analysis software described in 
\citet{sug11} and \citet{nak11}, and include only data obtained 
with the GSC counters operated with the nominal high voltage (1650 V).
We exclude events detected by two carbon anodes
\#1 and \#2 in all the counters whose responses have relatively large
calibration uncertainties at present \citep{sug11}. We extracted the 
source photons from a region of a circle centered at the target
position with a radius of 1.5 deg, and that for the background from a
circle with a radius of 3.3 deg excluding a region within 2.0 deg
from the target. The image of extracted data is presented
in Figure~\ref{maxi_img}, where the regions for source and background 
are overlaid. We remove the data obtained by a scan that did not cover
the whole source and background regions. It is confirmed that the energy 
response of the GSC is quite reliable within these selection criteria,
from an extensive calibration using the Crab nebula \citep{nak11}.
After the data screening, we accumulate the spectra every 3 days 
in MJD 55310--55344 and every week in MJD 55345--55550, 
considering the variability of the X-ray fluxes as shown in 
Figure~\ref{maxi_publc}.

\subsection{Swift/XRT observation and Data Reduction}
 
Swift performed three pointing observations of GX 339$-$4 on MJD 55352, 
55359, and 55366 (corresponding to 2010 June 5, 12, and 19). 
To check the consistency with the MAXI spectra, we also analyze
the Swift/XRT data. These observations were all performed in the
Windowed Timing (WT) mode (1-dimensional imaging) to avoid
pile-up. The net exposures were 1.2, 1.3, and 1.0 ksec, respectively.

We reduce the Swift/XRT archival data with the standard pipeline
script {\tt xrtpipeline} provided in the HEAsoft analysis package
version 6.10. The source events are extracted from a box region of
40 pixels (along DETX coordinate) $\times$ 30 pixels whose center is
located at the target position. A 80 pixel $\times$ 30 pixel box is
chosen for the background area. In spite of WT mode adopted in these
observations, the data still suffer from pile-up as reported by
\citet{yan11}. Thus, we exclude the events in the ''core'' region of
the source area, a rectangular region of 6 pixels $\times$ 30 pixels
centered at the target.

The spectra are extracted with XSELECT version 2.4b. We utilize the 
response matrix file, {\tt swxwt0to2s6\_200709\\01v012.rmf} in the
Swift Calibration Database (CALDB) provided on 2011 February 9. The
ancillary response files for the three epochs are created with the
tool {\tt xrtmkarf} by using the exposure files produced in the pipeline
processing. We add a 3\% systematic error to each spectral bin to 
include possible calibration uncertainties.

\section{Analysis and Results}

\begin{figure*}
  \begin{center}
    \FigureFile(80mm,80mm){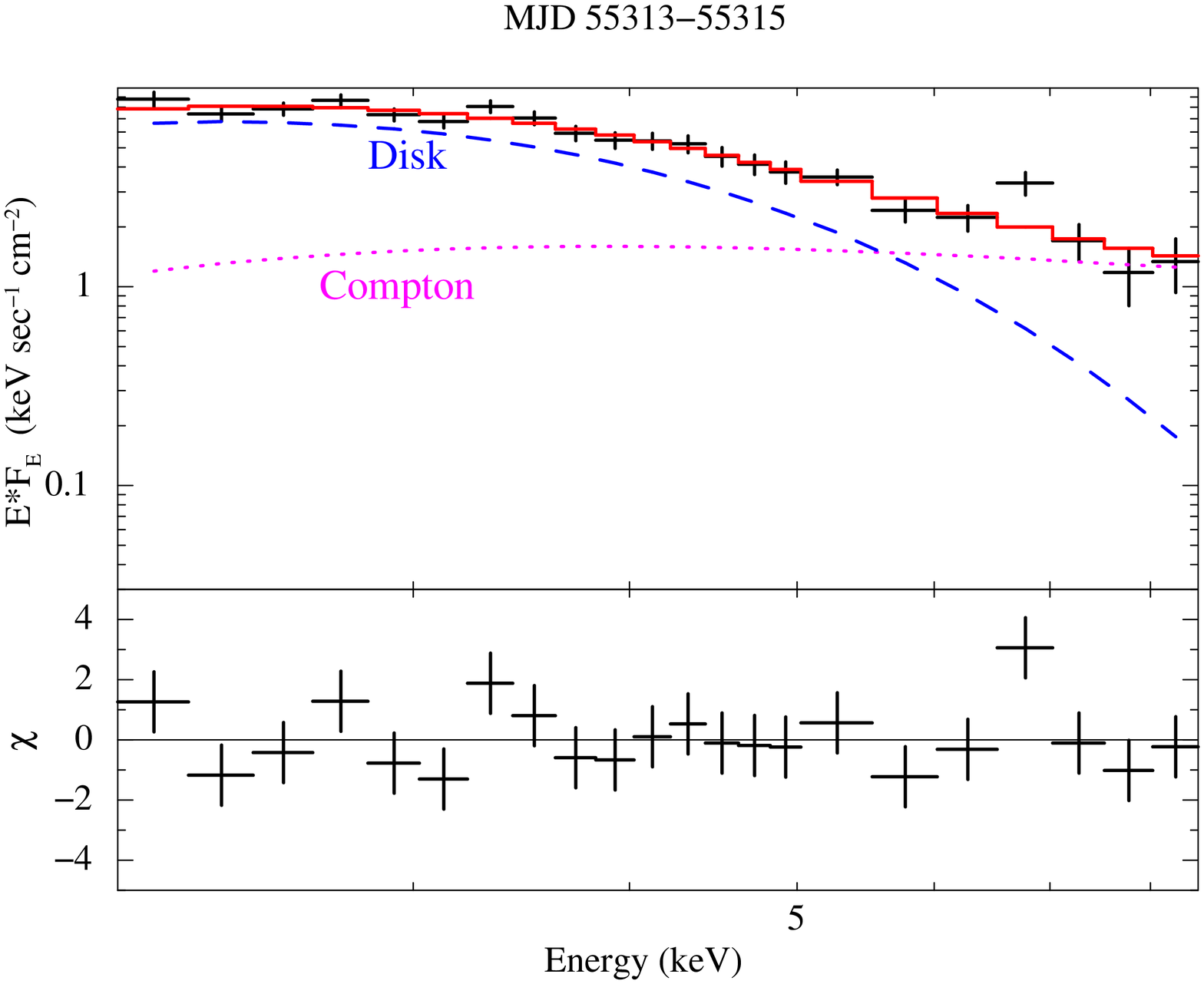}
    \FigureFile(80mm,80mm){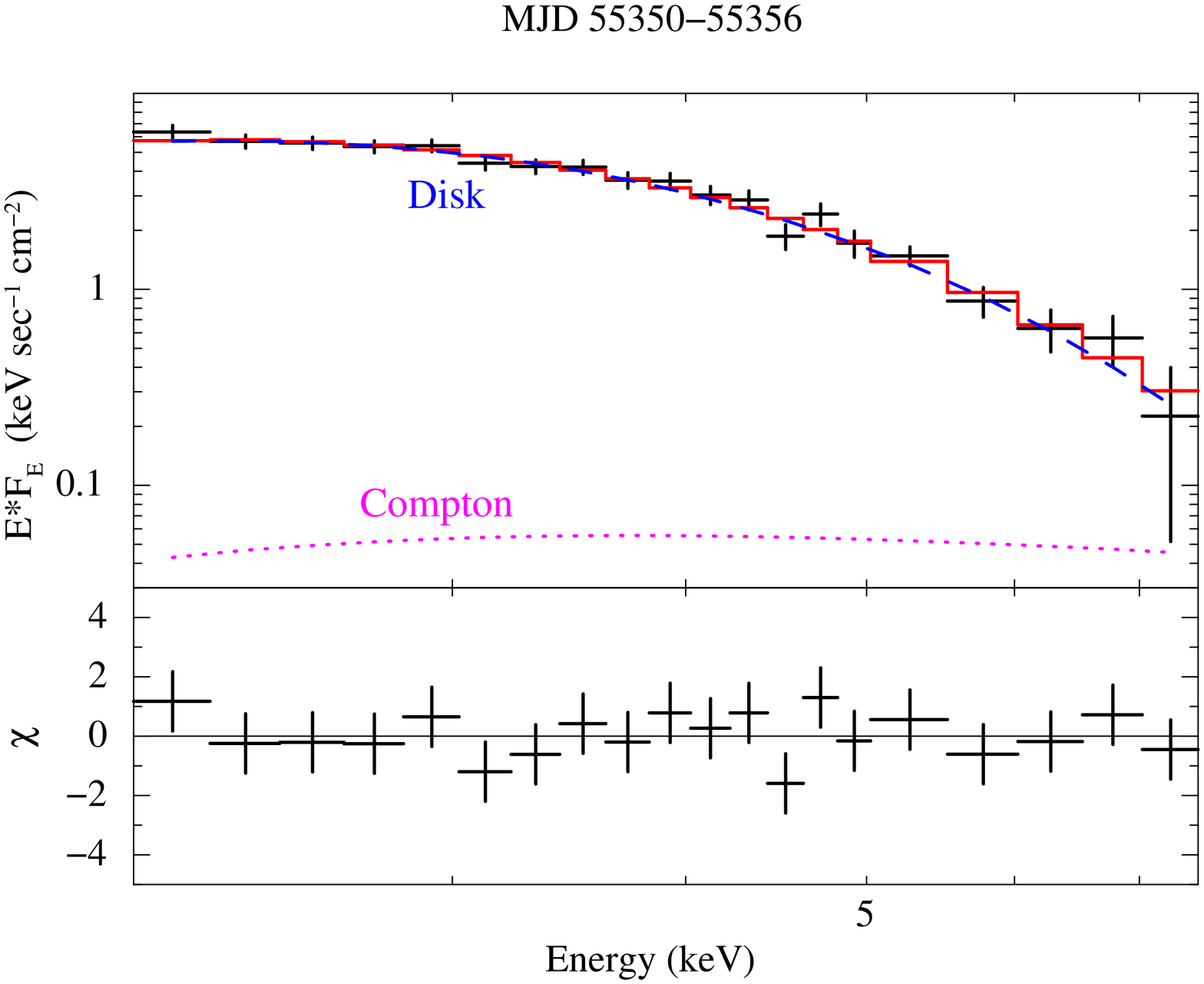}
    \FigureFile(80mm,80mm){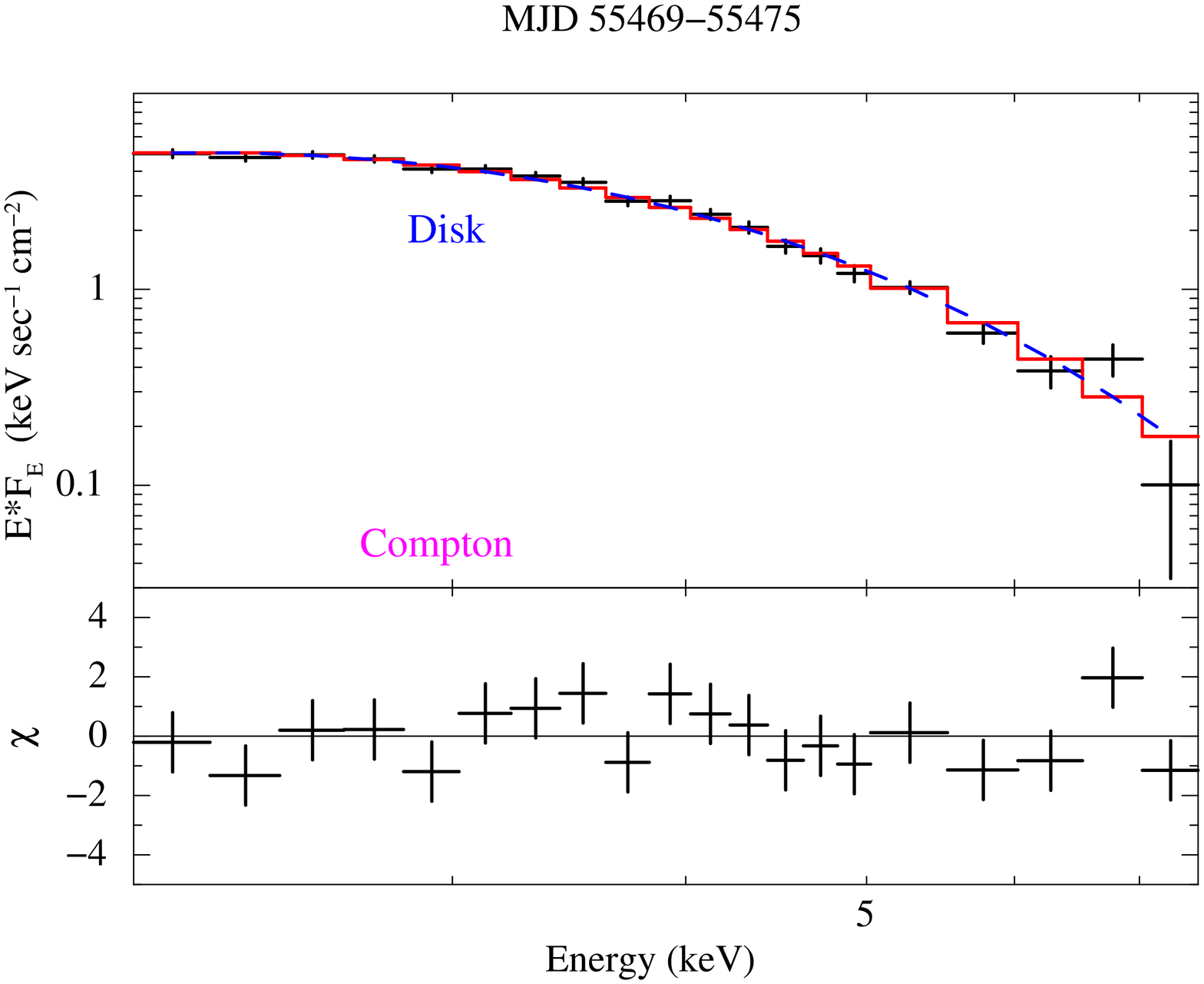}
    \FigureFile(80mm,80mm){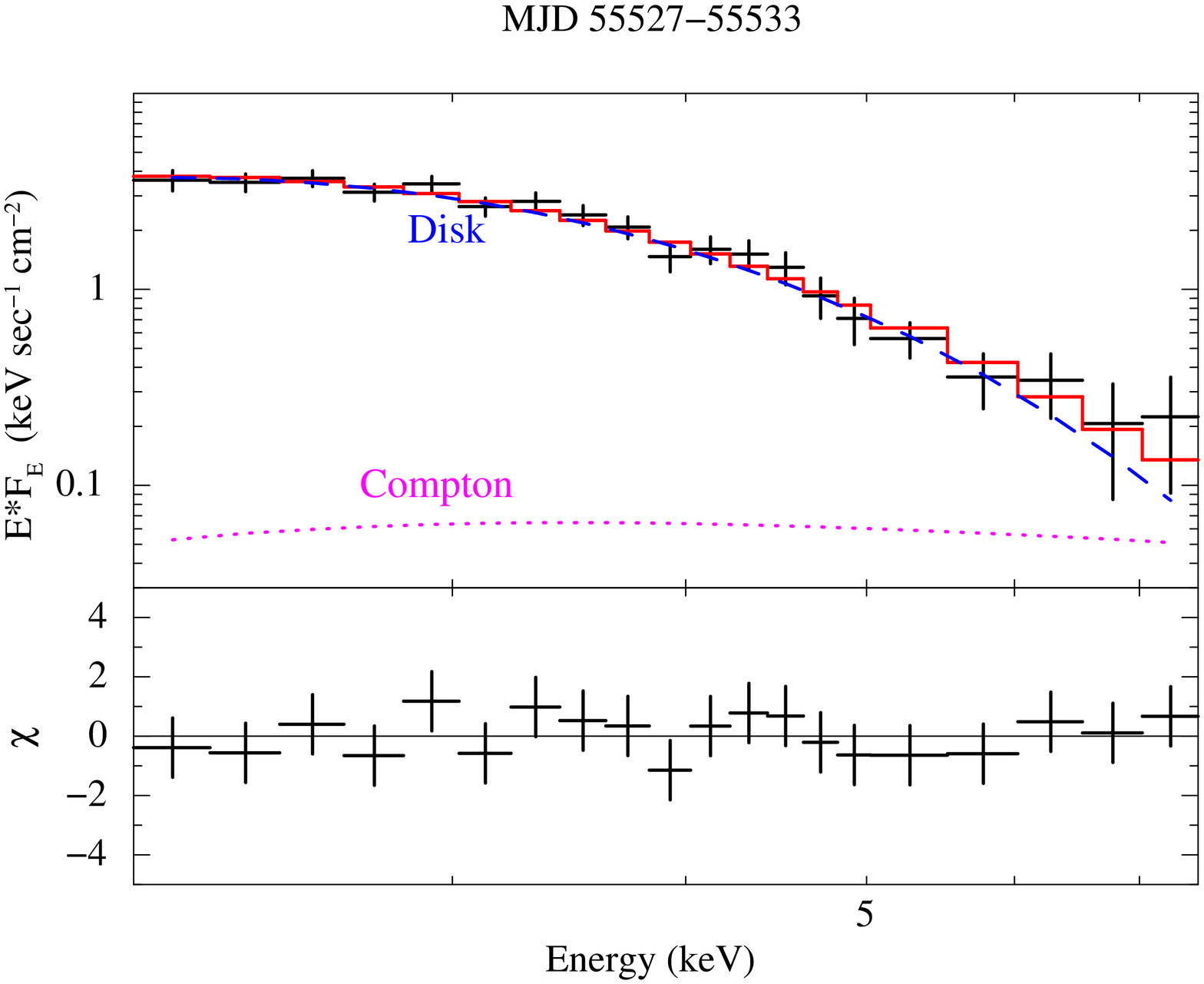}    
  \end{center}
  \caption{MAXI/GSC time-averaged spectra of GX 339$-$4 unfolded with
the response in the $\nu F_{\nu}$ form (crosses, black),
extracted from the data of MJD=55313--55315 (top left),
55350--55356 (top right), 55469--55475 (bottom left), and 55527--55533
(bottom right). The best-fit {\tt simpl*diskbb} model is
overplotted (solid, red). The contributions from the MCD (dashed,
blue) and its Comptonization components (dot, magenta) are separately
plotted. The residuals of the fit are shown in the lower panel in
units of $\chi$.
Notes. To plot the direct disk and Comptonized components
separately, we here represent the model with the form {\tt simpl*diskbb+diskbb}, 
where the scattered fraction of {\tt simpl} is
fixed at 1 and the normalization ratio of the first component to the
second one is set to the best-fit scattered fraction.
}\label{maxi_fit}

\end{figure*}

\begin{figure*}
  \begin{center}
    \FigureFile(80mm,80mm){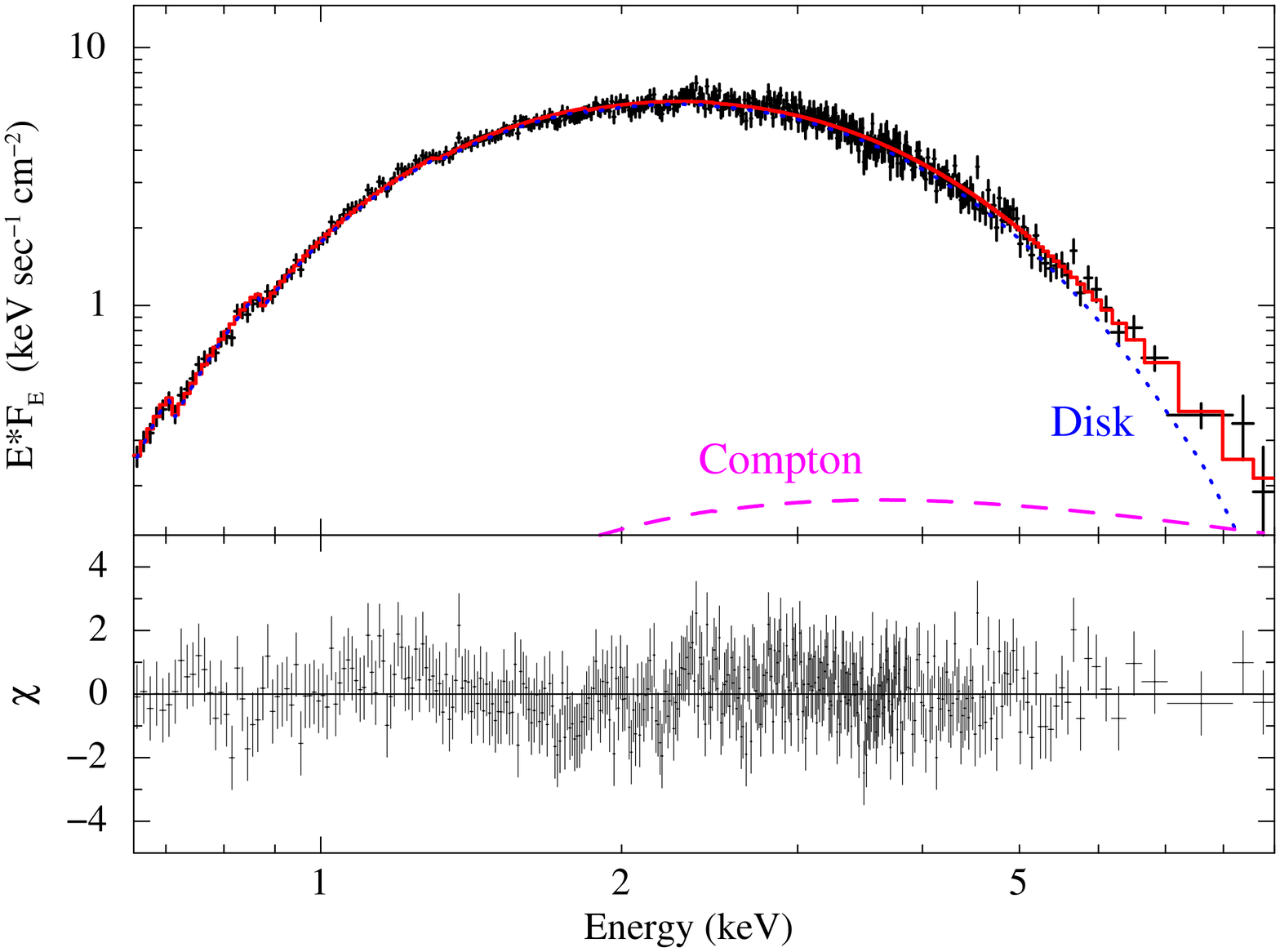}
    \FigureFile(80mm,80mm){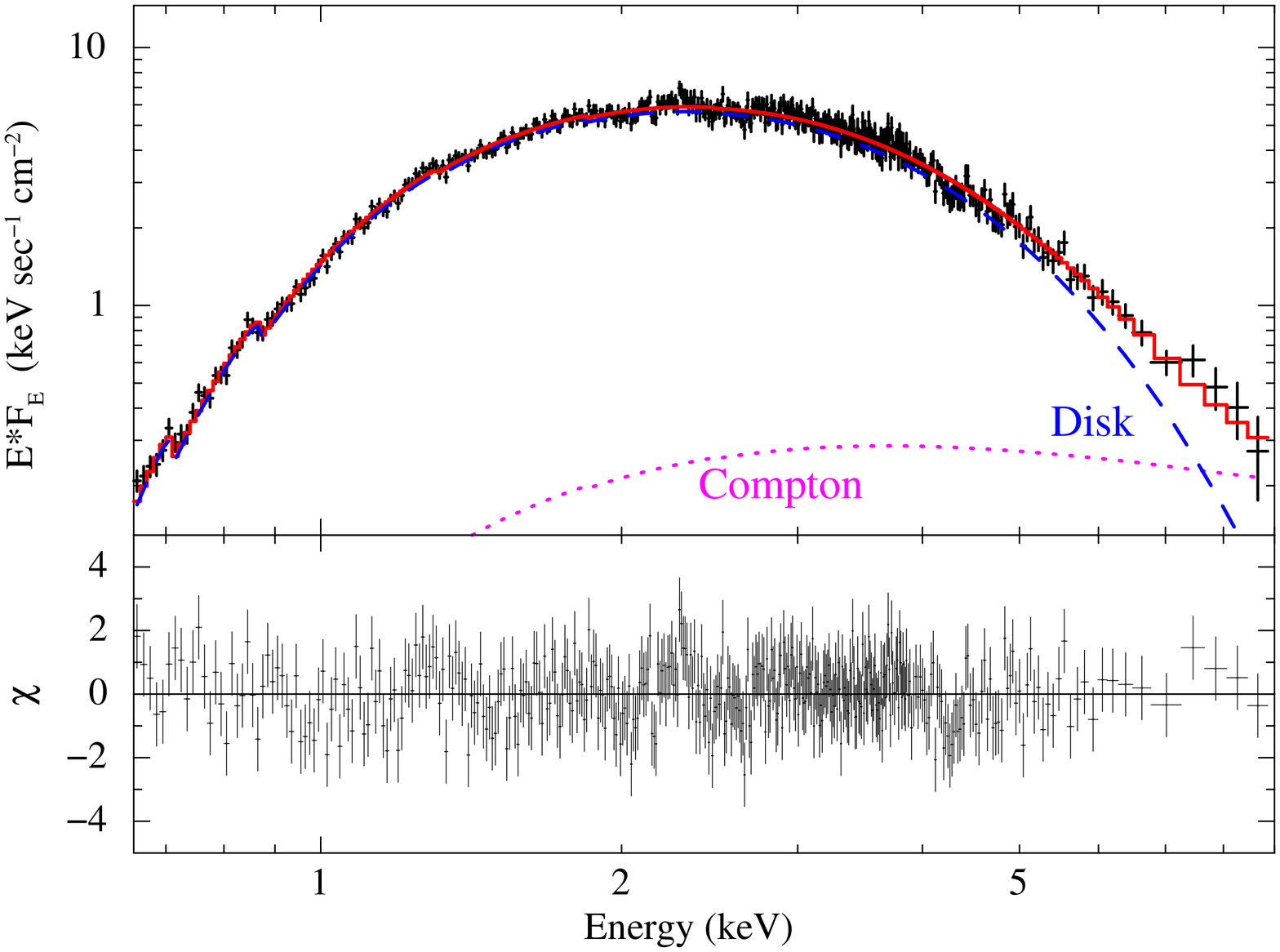}
    \FigureFile(80mm,80mm){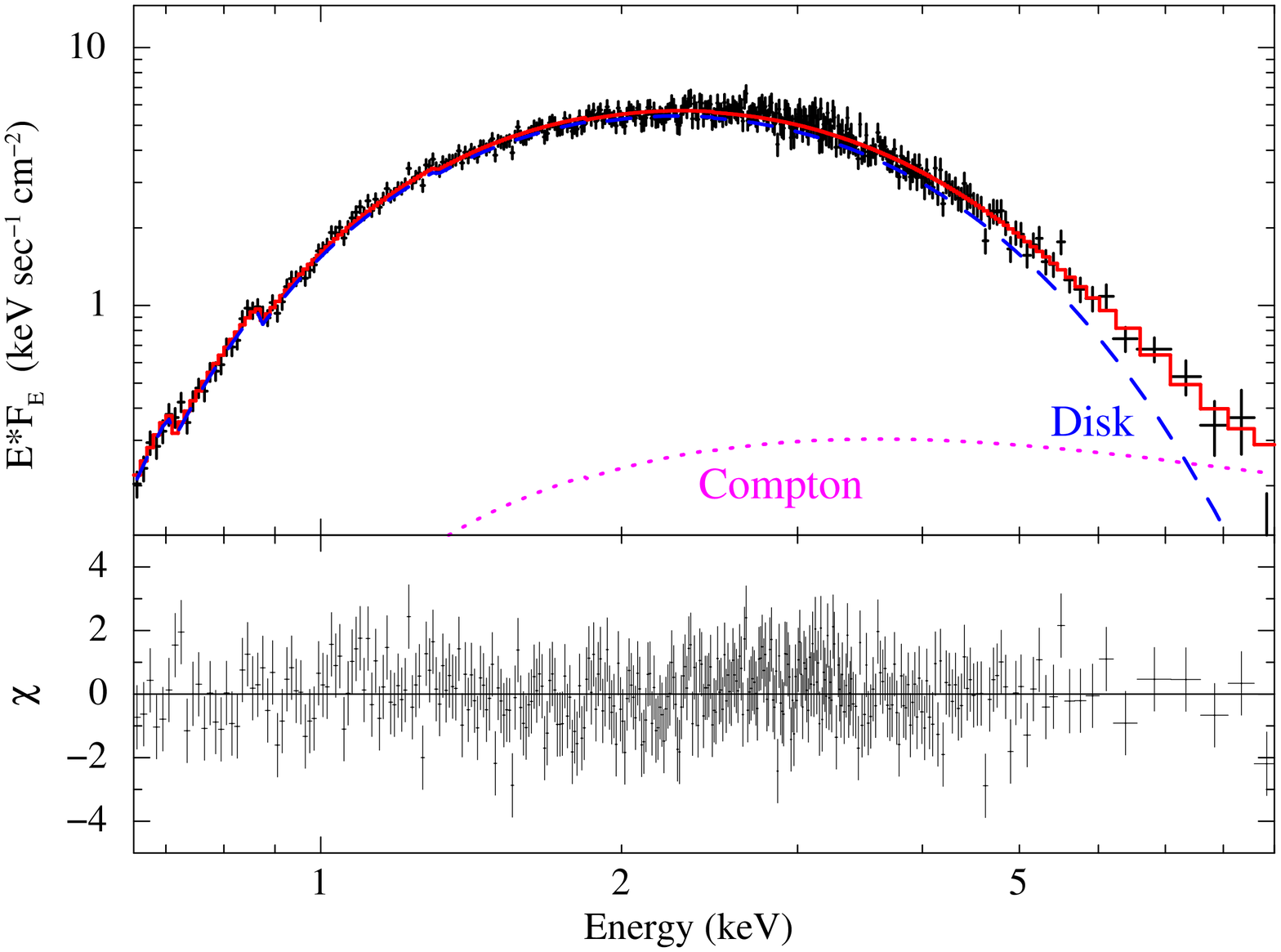}
  \end{center}
  \caption{
Swift/XRT spectra unfolded with the response in the $\nu
F_{\nu}$ form (crosses, black), observed on MJD=55352 (top left),
55359 (top right), and 55366 (bottom). The best-fit {\tt simpl*diskbb}
model is overplotted (solid, red) with separate contributions from the
MCD (dashed, blue) and its Comptonization component (dot magenta).
The residuals are shown in the lower panel in units of $\chi$.
}\label{swift_fit}
\end{figure*}

We uniformly fit all the MAXI/GSC and Swift/XRT
spectra to a partially Comptonized multi-color-disk blackbody (MCD) model
with an interstellar absorption. We adopt the {\tt diskbb} model
\citep{mit84} for the MCD and the {\tt simpl} model
\citep{ste09} for Compton scattering. The {\tt simpl} model is a
empirical convolution model that converts a given fraction of the
incident spectrum into a power law shape with a photon index
$\Gamma$. Because of the poor statistics in the hard X-ray ($>10$ keV)
band, we cannot constrain the photon index of the {\tt simpl} model.
Therefore, we fix it at $\Gamma = 2.5$, as a typical value for 
the hard tail of BHBs in the high/soft state \citep{mcc06}.
Fitting with a convolution model requires extending the energy range. 
We set it to 0.01--1000 keV. We use the {\tt phabs} model as the interstellar
absorption. The column density is fixed to $N_{\rm H}=0.5 \times 10^{22}$
cm$^{-2}$ (\cite{men97}; \cite{kon00}) for the fit of MAXI/GSC data,
while left free for that of Swift/XRT, since the XRT has a
sensitivity down to $\sim 0.4$ keV. The model is expressed as 
{\tt phabs$\times$(simpl*diskbb)} in the XSPEC
terminology. The free parameters of the continuum is three in total, the
innermost temperature and normalization of the MCD component, and the
Comptonized fraction in the {\tt simpl} model.

\begin{table*}
  \caption{Best-fit parameters of the MCD model.}\label{tab_MCD}
  \begin{center}
    \begin{tabular}{ccccccc}
      \hline
      instrument & MJD & $N_{\rm H}$& $T_{\rm in}$ & Compton & $R_{\rm in}$ & $\chi^2$ /d.o.f \\
       &  & ($10^{22}$ cm$^{-2}$) & (keV) & Fraction (\%) & (km) & \\
      \hline
      MAXI/GSC  & 55313--55315 & $0.5$ (fix) & $0.80^{+0.07}_{-0.08}$ & 17$^{+7}_{-6}$ 
& $67^{+13}_{-16}$ & 23/20 \\
      MAXI/GSC  & 55350--55356 & $0.5$ (fix) & $0.78^{+0.06}_{-0.04}$& $<5$ 
& $60^{+12}_{-6}$ & 11/20 \\
      MAXI/GSC  & 55469--55475 & $0.5$ (fix) & $0.75^{+0.01}_{-0.02}$ & $<1$ & $60^{+4}_{-3}$ & 19/20 \\
      MAXI/GSC  & 55527--55533 & $0.5$ (fix) & $0.70^{+0.05}_{-0.06}$ & $<5$ & $61^{+15}_{-10}$ & 9/17 \\
      \hline
      Swift/XRT & 55352 & $0.426^{+0.005}_{-0.007}$ & $0.806^{+0.008}_{-0.009}$ 
& $1.9^{+0.8}_{-0.7}$  & $57 \pm 1$ & 317/391  \\
      Swift/XRT & 55359 & $0.486 \pm 0.008$ & $0.812 \pm 0.009$ & $3.3^{+0.8}_{-0.9}$ 
& $55 \pm 1$ & 347/391 \\
      Swift/XRT & 55366 & $0.444^{+0.006}_{-0.008}$ & $0.796^{+0.009}_{-0.006}$ & $3.6^{+0.8}_{-0.9}$ 
& $56^{+1}_{-4}$ & 301/341  \\
      \hline
    \end{tabular}
  \end{center}
\end{table*}

We obtain acceptable fits from all the spectra in terms
of $\chi^2$. Figures~\ref{maxi_fit} and \ref{swift_fit}
display examples of the MAXI/GSC and Swift/XRT spectra, 
respectively, in the $\nu F_\nu$ form unfolded with the 
energy response. The best-fit model is overplotted on the 
data, and the residuals in units of $\chi$ are
shown in the lower panels. The resulting parameters are listed in
Table~\ref{tab_MCD}. The fraction of the Comptonized component is
found to be small ($\leq 25$\%) over the entire periods, 
indicating that GX 339--4 stayed in the high/soft state, not in the
intermediate (or very high) state characterized by a strong
Compton component. As shown in Table~\ref{tab_MCD}, the
absorption column density obtained from the Swift/XRT spectra, $\approx
0.45 \times 10^{22}$ cm$^{-2}$, is somewhat smaller than $0.5 \times
10^{22}$ cm$^{-2}$ adopted for the spectral fits to the MAXI/GSC
spectra, but we confirm that the best-fit inner disk radius derived from
the MAXI/GSC data are unchanged ($<1\%$) within the 90\% confidence
errors even when we fix $N_{\rm H}=0.4 \times 10^{22}$ cm$^{-2}$.

\begin{figure}
  \begin{center}
    \FigureFile(80mm,80mm){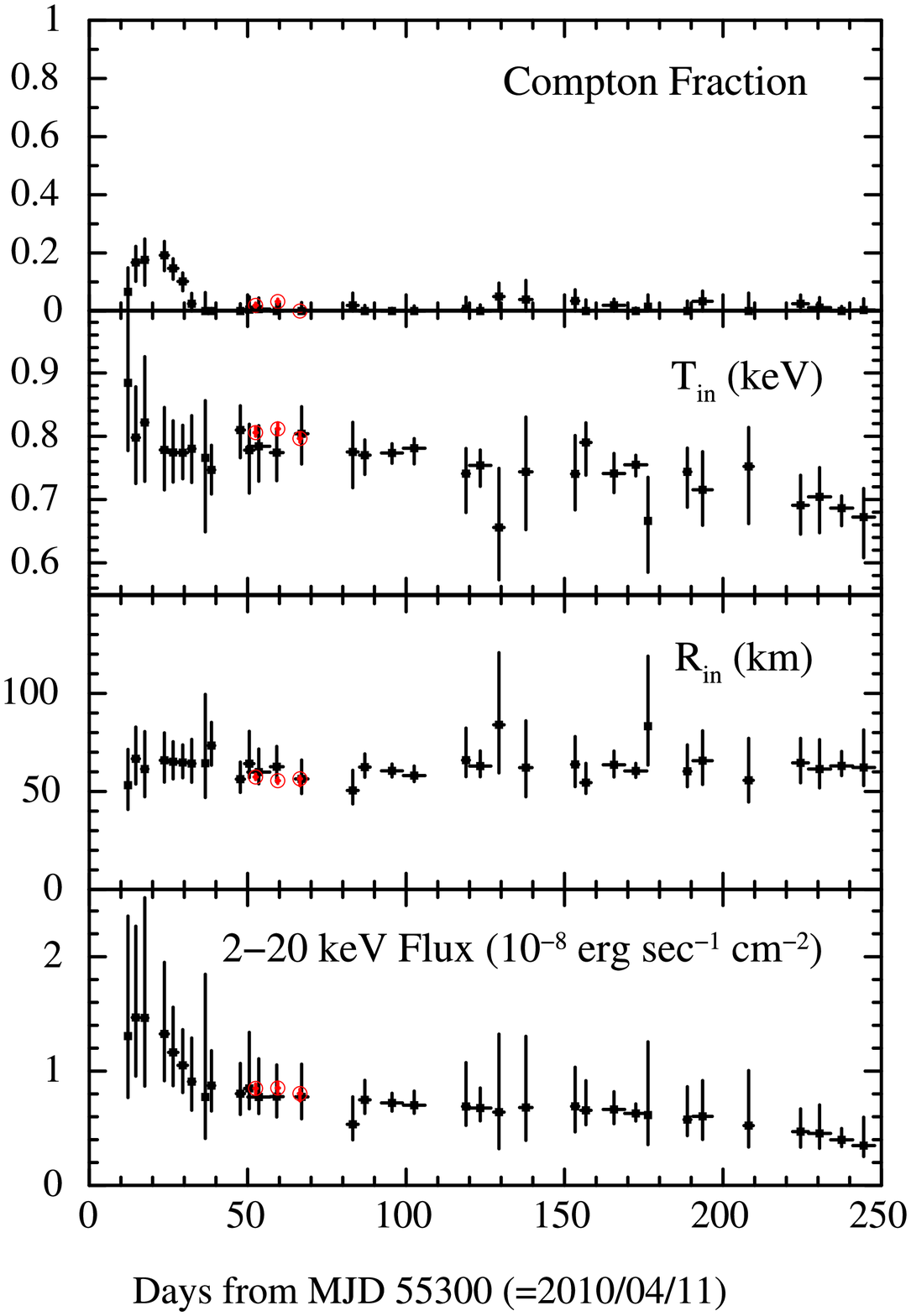}
  \end{center}
 \caption{
Evolution of Comptonized fraction in the {\tt simpl} model (top),
innermost temperature $T_{\rm in}$ (second), inner disk radius $R_{\rm
in}$ estimated from the normalization of {\tt diskbb} component
(third), and unabsorbed 2--20 keV flux (bottom).  
The MAXI and Swift results are plotted in black (filled square) 
and red (open circle),
respectively.}\label{date-par}
\end{figure}

\begin{figure}
  \begin{center}
    \FigureFile(80mm,80mm){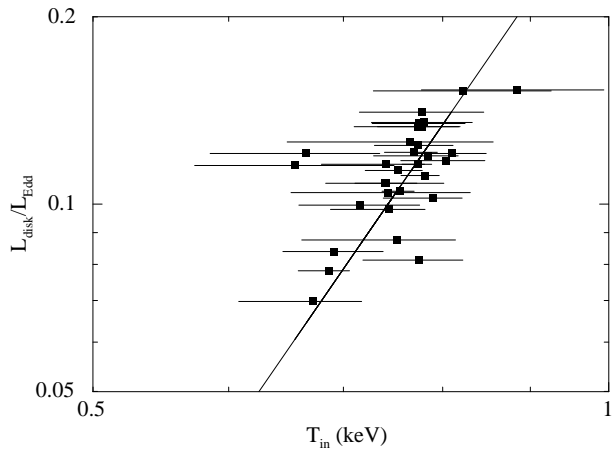}
  \end{center}
 \caption{
The luminosity-temperature relation of the disk component of GX 339--4
in the high/soft obtained from the MAXI/GSC data with the MCD fit. 
The data points are fitted by a power-law with $L_{\rm disk} \propto 
T_{\rm in}^4$ (solid line). 
}\label{Ldisk_Tin}
\end{figure}

Figure~\ref{date-par} plots the evolution of the Comptonized fraction
(top panel), innermost temperature of the MCD component (second), inner
disk radius (third), and 2--20 keV band flux (bottom). 
The innermost radius is derived from the
normalization of the MCD component, by applying the combined
correction factor of 1.18 for the stress-free boundary condition 
($\xi=0.41$; see \cite{kub98}) and the color hardening factor
($\kappa=1.7$, \cite{shi95}).
Here we adopt the inclination angle $i = 46^{\circ}$ from
the \citet{shi11}, and assume the distance $d=$8 kpc as the most reasonable
estimate \citep{zdz04}.
It can be noticed that the inner radius stayed almost constant at $\sim 60$~km, 
while the temperature gradually decreased with time according to the flux change.
The constancy of the radius is consistent with the generally accepted
picture of the high/soft state in which the standard disk is always
extending to the ISCO.
Figure~\ref{Ldisk_Tin} plots the disk bolometric luminosity $L_{\rm disk}$ 
in terms of Eddington luminosity $L_{\rm Edd}$ versus the inner temperature 
$T_{\rm in}$ obtained from the MCD fit of the MAXI/GSC data. We assume the 
distance, inclination, and black hole mass to be $d=8$ kpc, $i=46^\circ$, 
and $6.8 M_\odot$, respectively. As shown in Figure~\ref{Ldisk_Tin}, the 
data points roughly follow a simple relation $L_{\rm disk} \propto T_{\rm in}^4$,  
which is expected from MCD emission with a constant innermost radius.

To check the effects on the estimated inner disk radius by
the assumed geometry of the Comptonizing corona,
we also calculate $R_{\rm in}$ from the MAXI/GSC spectra on MJD 
55325--55327 as representative data, assuming that the Comptonized 
emission modelled by {\tt simpl} is isotropic.
We use the equation of photon conservation expressed as follows
\citep{kub04}:
\begin{eqnarray}
F^p_{\rm disk}+F^p_{\rm thc}2 \cos i &= 0.0165 \left[ \frac{r_{\rm in}^2\cos i}
{(D/10\mbox{ kpc})^2}\right] 
\left( \frac{T_{\rm in}}{1\mbox{ keV}} \right)^3 \nonumber \\ 
 & \mbox{ photons } {\rm s}^{-1} \mbox{ }{\rm cm}^{-2}. \label{eq1}
\end{eqnarray}
where $F^p_{\rm disk}$ and $F^p_{\rm thc}$ are the unabsorbed
0.01--100 keV photon flux from the disk and Comptonized components,
respectively. 
Taking into account the correction factor of 1.18
\citep{kub98}, we obtain the inner disk radius of $R_{\rm in} =
64^{+6}_{-5}$ km. 
It is consistent with that calculated from the normalization of 
the MCD component (as plotted in Figure \ref{date-par})
by assuming disk-like geometry for the Comptonized emission.
Also, we find that these results are not changed when we replace {\tt
simpl} with {\tt nthcomp}, a more physical model for thermal
Comptonization.  Hence, we can ignore the systematic uncertainties due
to Comptonization in discussing the inner disk radius.

\begin{table*}
  \caption{Best-fit parameters of relativistic disk models from the Swift/XRT data on MJD 55359.}\label{tab_kerr}
  \begin{center}
    \begin{tabular}{cccccccccc}
      \hline
       model & $N_{\rm H}$ & $a$ & $i$ & $d$ & $M_{\rm BH}$ & $\dot{M}$ & $L_{\rm disk}/L_{\rm Edd}$ & Compton & $\chi^2$/d.o.f \\
        & ($10^{22}$ cm$^{-2}$) & & (deg) & (kpc) & ($M_\odot$) & ($10^{18}$ g s$^{-1}$) &   & Fraction (\%) & \\
      \hline
       {\tt kerrbb}\footnotemark[$*$] & $0.509^{+0.009}_{-0.008} $ & $ <0.05 $ & $38^{+2}_{-0} $ & 8 (fix) 
& $4.6^{+0.2}_{-0.1}$ & $2.27^{+0.05}_{-0.03}$ &  & $ 1.9^{+0.5}_{-0.8} $ & 363/389  \\
       {\tt bhspec}\footnotemark[$\dagger$] & $0.530 \pm0.08 $ & $ <0.33 $ & $38^{+5}_{-0}$ & 8 (fix) 
& $4.6^{+0.5}_{-0.1}$ & & $0.226^{+0.004}_{-0.017}$ & $ 3.5^{+0.5}_{-0.6} $ & 345/389  \\
      \hline
 \multicolumn{10}{@{}l@{}}{\hbox to 0pt{\parbox{180mm}{\footnotesize
       \par\noindent
       \footnotemark[$*$] Self-irradiation is not included (rflag=0), and limb-darkening is taken into account (lflag=1).
       The torque-free inner edge is assumed.
       \par\noindent
       \footnotemark[$\dagger$] We fixed $\log(\alpha)=0.01$, where $\alpha$ is the viscosity parameter 
        in the \citet{sha73} prescription for the stress $\tau_{r \phi}=\alpha \times P$ ($P$ is the total pressure).
     }\hss}}
    \end{tabular}
  \end{center}
\end{table*}

\begin{figure*}
  \begin{center}
    \FigureFile(80mm,80mm){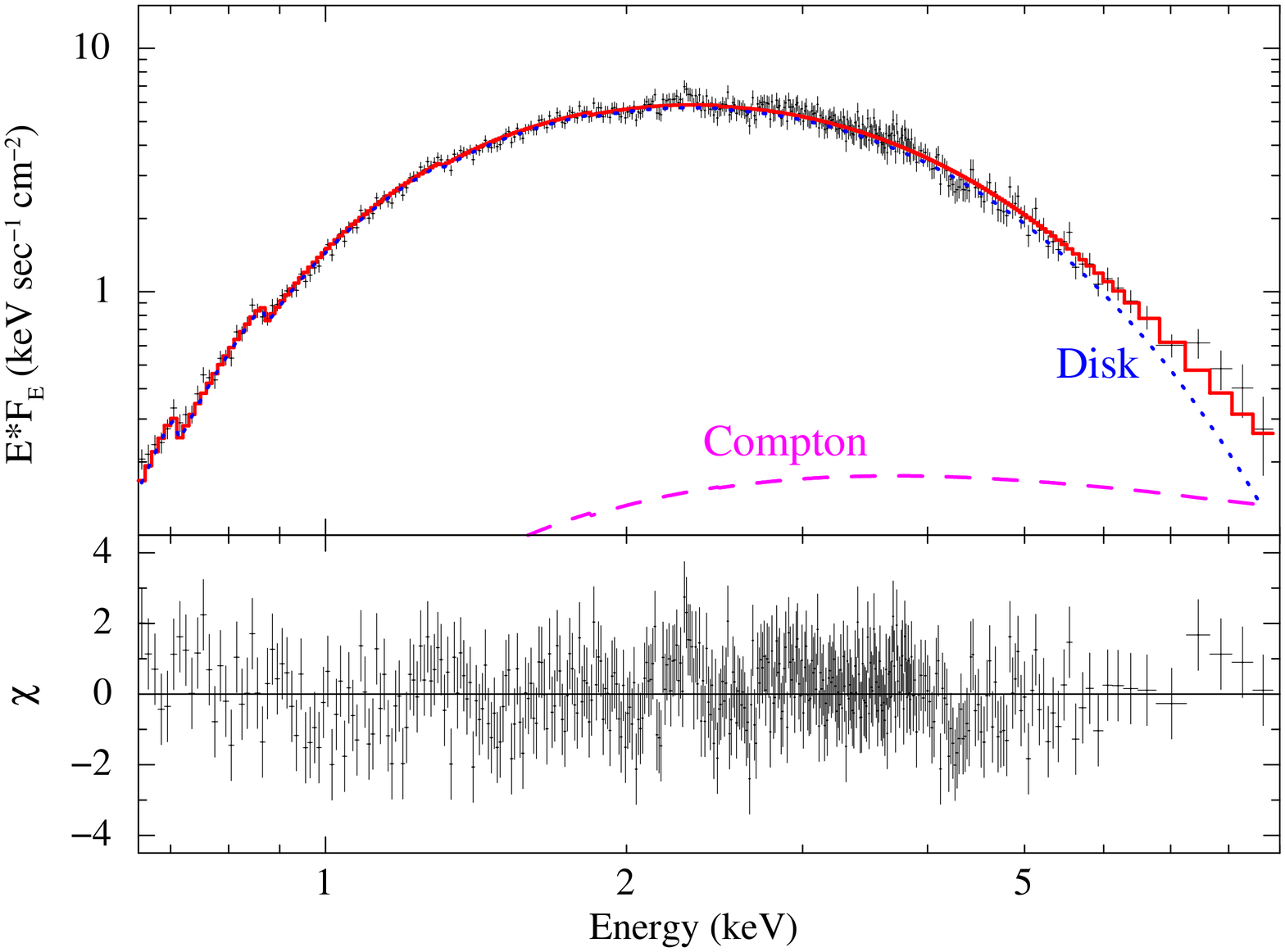}
    \FigureFile(80mm,80mm){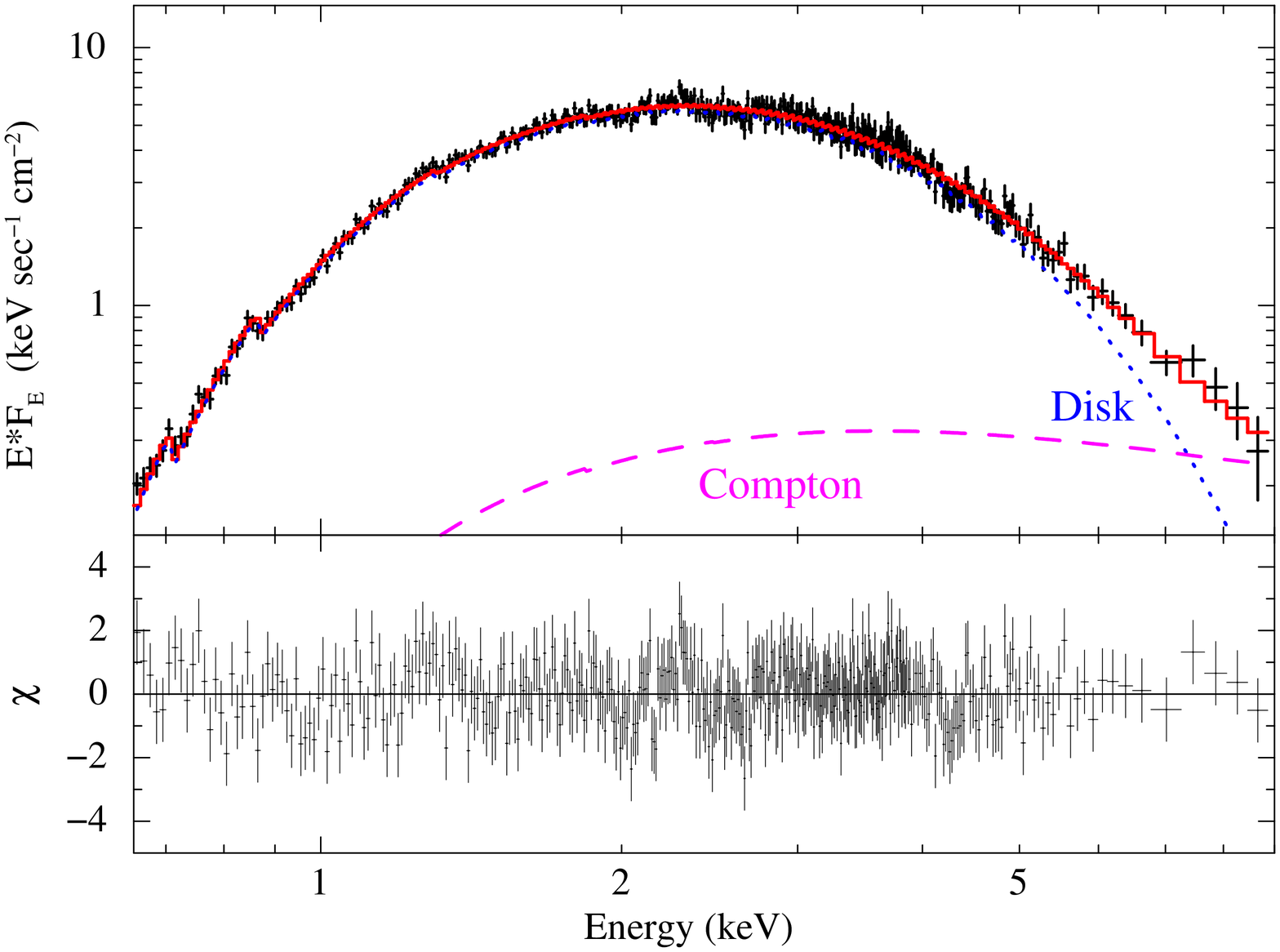}
  \end{center}
  \caption{
Swift/XRT spectrum on MJD=55359 unfolded with the response in the $\nu
F_{\nu}$ form (crosses, black). The best-fit {\tt simpl*kerrbb} (left)
and {\tt simpl*bhspec} (right) models are overplotted (solid, red)
with separate contributions from the disk emission (dashed, blue) and
its Comptonization component (dot magenta). The residuals are shown
in the lower panel in units of $\chi$.
}\label{swift_fit_kerr}
\end{figure*}

The real disk spectrum around a black hole, however, should deviate
from the MCD spectrum due to the inner boundary condition, and to
relativistic effects including gravitational redshift, beaming, and
light bending, which become particularly important for a rapidly
spinning black hole. Precise measurement of the spectral shape of the
disk emission may be used to obtain information on the spin of a black
hole even if all system parameters, the black hole mass, distance, and
inclination, are not accurately known. With an aim to constrain the spin
parameter of GX 339--4, we analyze the Swift/XRT data with the more
advanced disk models {\tt kerrbb} \citep{li05} and {\tt bhspec}
\citep{dav05} instead of {\tt diskbb}. Both codes relativistically
compute the spectrum from a geometrically-thin accretion disk around a
spinning black hole with an arbitrary spin parameter $a \equiv cJ/GM^2$,
where $M$ and $J$ are the mass and angular momentum of the black hole,
$c$ the speed of light, and $G$ the gravitational constant
\citep{sha86}. Similarly to the previous MCD fit, the disk emission is
convolved with {\tt simpl} to represent the Comptonized component.

We fit the Swift/XRT spectrum observed on MJD 55359 as representative
data with the model expressed as {\tt phabs}$\times$({\tt simpl}$*${\tt
kerrbb}) or {\tt phabs}$\times$({\tt simpl}$*${\tt bhspec}). The
distance is fixed at $d=8$ kpc, while the spin parameter and 
mass accretion rate ({\tt kerrbb}) or Eddington ratio ({\tt bhspec})
are left as free parameters.
We assume $a \geq 0$ (i.e., the spin direction
of the black hole is the same as that of the accretion disk). In the
{\tt kerrbb} model, the spectral hardening factor is fixed at 1.7, and
zero torque is assumed at the inner edge of the disk. 
First, we fix the inclination at $i=46^\circ$ (the best estimate of
\cite{shi11}), and the black hole mass at $M=15.6 M_\odot$ calculated
from the mass function of $f(M)=5.8 M_\odot$ \citep{hyn03}. 
The fit is acceptable ($\chi^2/\nu = 395/391$), and the spin parameter
is constrained to be $a=0.94^{+0.02}_{-0.01}$ (in the case of the {\tt kerrbb}
model). Next, considering the uncertainty in the mass function
($f(M)>2 M_\odot$), we make both black hole mass and inclination free
parameters. The inclination is allowed to vary only in the range of
$38^\circ \leq i \leq 54^\circ$ \citep{shi11}. The fitting results
are shown in Table \ref{tab_kerr} and Figure \ref{swift_fit_kerr}. The
fit is significantly improved ($\chi^2/\nu = 363/389$) compared with
the previous fit where $M$ and $i$ are fixed.  We obtain the spin
parameters $a< 0.05$ or $a< 0.33$ with the black hole mass of 
$M = 4.6^{+0.2}_{-0.1} M_\odot$ or $M = 4.6^{+0.5}_{-0.1} M_\odot$ from 
the {\tt kerrbb} or {\tt bhspec} fit, respectively.
This result that a small value of the $a$ parameter is favored is not
changed even when the energy band below 1 keV is ignored to avoid any
systematic uncertainties in the low energy response, or when a smaller
photon index of $\Gamma =2.0$ is adopted for the {\tt simpl} model
instead of $\Gamma=2.5$. The same trend is also confirmed from
the other two Swift/XRT spectra taken on MJD=55352 and 55366.

\section{Discussion and Conclusion}

We have acquired uniform-quality X-ray spectra of GX 339--4 every 3 or 7
days in the high/soft state during the 2010 outburst from the MAXI/GSC
monitoring data. All the spectra are found to be well represented by a
partially Comptonized MCD model, from which we are able to estimate the
inner disk radius. As shown in Figure~\ref{date-par}, the inner radii
obtained from the MAXI/GSC data are consistent with those of
the simultaneous Swift/XRT data, which covers a softer energy band in the 
0.6--10 keV and is more sensitive to determine the parameters of
the MCD component. The consistency between MAXI/GSC and Swift/XRT indicates 
that our MAXI results are quite reliable, not subject to systematic 
uncertainties in the responses. The inner radius stayed almost 
constant over the whole period of 8 months, 
during which the X-ray flux changed by a factor of 3. This confirms 
the standard model scenario in the high/soft state that the 
accretion disk is (stably) extending down to the ISCO.

Our best estimate of the inner disk radius is 
\begin{equation}
\label{eq2}
R_{\rm in} = (61 \pm 2) \left( \frac{d}{8 {\rm kpc}} \right) 
\left( \frac{{\rm cos}(46^\circ)}{{\rm cos}(i)} \right)^{\frac{1}{2}} \ {\rm km},
\end{equation}
which is a weighted average from the MAXI/GSC results over the period from
MJD 55310 to MJD 55550.  We confirm that this value is consistent with previous
measurements using the MCD flux in the high/soft state. In 1983,
\citet{mak86} observed GX 339$-$4 in outburst three times with the
Tenma satellite.  We find that their estimated inner radius becomes
$59 \pm 1$, $59 \pm 1$, and $58 \pm 1 $ km for $i=46^\circ$ and $d=8$
kpc, after correcting for the boundary condition and color hardening
factor as described above. \citet{bel99} performed three
target-of-opportunity observations of GX 339--4 in the 1998 outburst
with RXTE, two of which correspond to the period of the high/soft
state. They obtained the inner radius of $80 \pm 11$ and $72 \pm 1$ km
for $i=46^\circ$ and $d=8$ kpc. These values are somewhat larger than
the Tenma result (\cite{mak86}) as well as our MAXI/GSC and Swift/XRT
results. The reason is yet unclear and we do not pursue this in our
paper.

Assuming that the stable inner disk of a BHB in the high/soft state
corresponds to the ISCO, we can estimate the black hole mass from the
obtained $R_{\rm in}$ value, which is coupled with the spin parameter
$a$. For a
non-spinning black hole, the ISCO correspond to $6 R_{\rm g}$ ($R_{\rm
g}$ represents the gravitational radius $GM/c^2$). Using our best
estimate on the inner disk radius from the MAXI data, $61 \pm 2$ (km),
we constrain the black hole mass of GX 339$-$4 to be $6.8 \pm 0.2
M_{\odot}$ by assuming $i=46^\circ$ and $d=8$ kpc in the case of
non-spinning black hole. 
This value is consistent with the constraint of
$M=4-16 M_{\odot}$ discussed in Section~4 of \citet{shi11} based on the
comparison of the iron-K line profile and the MCD component from the
Suzaku spectra in the low/hard state.

\begin{figure}
  \begin{center}
    \FigureFile(80mm,80mm){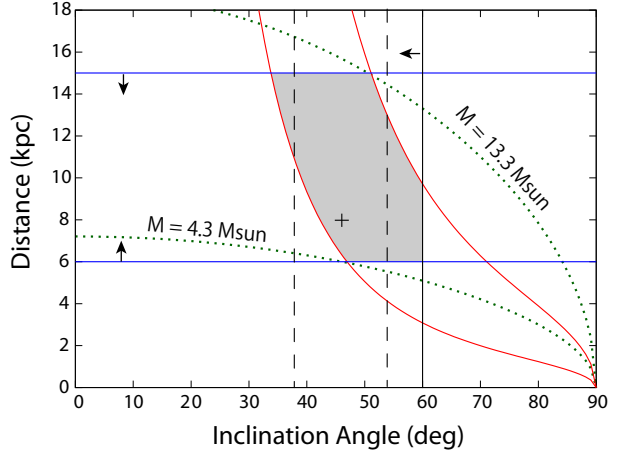}
  \end{center}
 \caption{
Diagram of constraints on the distance and the inclination
angle of GX 339--4. The two solid curves (red) are obtained from our
estimate of the inner disk radius $R_{\rm in}$ and the mass function
$f(M) = (2.0-6.3)M_\odot$. The spin parameter of $a=0$ is
assumed. The lower and upper limits on the distance (blue) are taken
from by \citet{hyn04}.  The upper limit of the inclination ($i \leq
60^\circ$) from the absence of an eclipse \citep{cow02} and its 90\%
confidence region obtained by \citet{shi11} from the iron-K line
profile are plotted by the solid and dashed lines, respectively. The
cross denotes the reference point used for our best estimate of the
black hole mass. The constant black-hole mass curves are plotted by
dotted green lines for the case of $M=13.3 M_{\odot}$ (upper) and
$M=4.3 M_{\odot}$ (lower).  }\label{incl_d}
\end{figure}

Combined with the mass function $f(M)$, defined as 
\begin{equation}
f(M) =
\frac{M}{\left(1+\frac{M_{\rm c}}{M} \right)^2} \sin^3 i \approx M
\sin^3 i,
\end{equation} 
where $M_{\rm c}$ represents the mass of the companion
star, the radius of the ISCO, $R_{\rm in} = 61\pm2$ (km),
constrains the relation between the inclination and distance for an
assumed spin parameter $a$. In the approximation, we ignore the
term $\frac{M_{\rm c}}{M}$, which is less than 0.08 \citep{zdz04}.
The $f(M)$ value of GX 339--4 was estimated by \citet{hyn03} to be
$(5.8\pm0.5) M_\odot$ (2.0 $M_\odot$ at a 95\% confidence lower limit), which
can be converted to a black hole mass (hence the ISCO for an assumed
spin) at a given inclination angle. By equating it with the
equation~(\ref{eq2}), we can determine the distance. The gray region
in Figure~\ref{incl_d} shows the constraints on the inclination and
distance obtained in this way within the allowed range of $f(M)=
(2.0-6.3) M_\odot$, assuming $a=0$ (i.e., $R_{\rm in} = 6 R_{\rm
  g}$). Here only the $i<60^\circ$ region is considered, which comes
from the absence of an eclipse \citep{cow02}, and the distance is
limited to be 6 kpc $<d<$ 15 kpc from the structure of
Na\emissiontype{D} line \citep{hyn04}.  A relation between the
inclination and distance corresponding to a given black hole mass can
be derived from the equation~(\ref{eq2}), where $a=0$ is assumed. As
shown in dotted green lines in Figure~\ref{incl_d}, the allowed upper
and lower limits of the black hole mass in the grey area obtained from
this relation are 13.3 $M_\odot$ and 4.3 $M_\odot$, which happens at
the intersections between the $d=$15 kpc line and $f(M)=6.3 M_\odot$
curve, and that between the $d=$6 kpc line and the $f(M)=2.0 M_\odot$
one, respectively.
We also
plot the constraint of $i=(46 \pm 8)^\circ$ (within dashed lines) obtained
by \citet{shi11} from the analysis of the iron-K line profile.
Thus, the allowed region in the $i$ and $d$ space derived from $R_{\rm
in}$ and $f(M)$ for the case of $a=0$ is well compatible with all these
previous constraints, including the point of $i=46^\circ$ and $d=8$
kpc as the best estimate.

For more detailed discussion, we need relativistic disk models
since the MCD model is only a simple approximation of the disk spectrum
(see e.g., \cite{kub10}).
The analysis of high quality X-ray spectra of a BHB in the high/soft state
with such models gives us information on the spin parameter of the
black hole.
Using the Swift/XRT spectrum, we obtain a large spin parameter 
$a\approx0.94$ when the mass and the inclination are fixed at $M=15.6
M_\odot$ at $d=8$ kpc and $i=46^\circ$, respectively. However, the resultant 
spin strongly depends on the assumed mass and inclination;
\citet{kol11} derived $a=0.1-0.5$ from the XMM-Newton
and RXTE spectra of GX 339--4 in the high/soft state with the {\tt bhspec}
model, assuming $10 M_\odot$ at
$d=8$ kpc with an inclination of $60^\circ$.
Thus, tighter determination of the system parameters of
GX 339--4 than that available at present is necessary
to derive a reliable answer.

We have also shown the possibility to constrain the black hole spin
from the X-ray continuum emission alone without independent
information on the mass. In fact, we find that the quality of the fit
is significantly improved when the mass and inclination are left as
free parameters, which favors a small value of the spin parameter
($a<0.05$ and $a<0.33$ with the {\tt kerrbb} and {\tt bhspec} models,
respectively). This can be understood because the relativistic
broadening of the disk spectrum in GX 339--4 looks less significant
than that predicted from a large spin parameter. 
It must be noted, however, that there remain systematic
uncertainties at $\sim$5\% level in the current disk models as
discussed by \citet{kol11}, which could affect the spectral fitting.
Further development of theoretical models reproducing the disk
emission would be very useful in this context.
Finally, we note that the best-fit black hole mass derived from the
{\tt kerrbb} and {\tt bhspec} fits are both $M \approx 4.6 M_\odot$ at
$d=8$ kpc with the spin parameter $a\simeq0$ and inclination
$i=38^\circ$. The mass is $\sim $30\% smaller than that obtained from the
same Swift/XRT spectrum with the MCD fit by assuming zero-spin ($M =
6.8 M_\odot$ for $d=8$ kpc and $i=38^\circ$). This difference should
be considered as a systematic error in estimating the black hole mass, 
which we must bear in mind in the critical 
discussion with accuracy better than $30$\% levels.

\bigskip


This work made use of the Swift public data archive.  This research
was partially supported by the Ministry of Education, Culture, Sports,
Science and Technology (MEXT), Grant-in-Aid No.19047001, 20041008,
20244015, 20540237, 21340043, 21740140, 22740120, 23000004, 23540265,
and Global-COE from MEXT ``The Next Generation of Physics, Spun from
Universality and Emergence'' and ``Nanoscience and Quantum Physics''.


\end{document}